\documentclass[conference]{IEEEtran}
\usepackage{graphicx,epsfig}

\begin{document}

\title{MAGIC multiwavelength observations:\\policy, and some recent results}

\author{\authorblockN{Alessandro De Angelis}
\authorblockA{Universit\`a di Udine, INFN and INAF, Italy,
and IST, Lisboa, Portugal\\
on behalf of the MAGIC Collaboration\\
Email: alessandro.de.angelis@cern.ch}
}
\maketitle

\begin{abstract}
MAGIC, 17 meters of diameter,  is  the world's largest single dish Imaging
Atmospheric Cherenkov Telescope, and reaches in the analysis the lowest energy threshold (60 GeV) among the VHE gamma detectors. Completed in September 2004, MAGIC started full operation with its first cycle of data taking in February 2005.
MAGIC observations in the galaxy cover, among
others, supernova remnants, the Galactic Center and binary systems. The low threshold makes of MAGIC the IACT
looking deepest in the Universe: the record of 
extragalactic sources detected includes Active Galactic Nuclei (AGN) at $z > 0.2$. Here we discuss the present performance of MAGIC
and the policy for the use of MAGIC data in multiwavelength campaigns. After a review of some recent highlights from MW studies, including the discovery of the most distant source ever detected (the AGN 3C279 at $z \simeq 0.54$), 
we present the expected performance of MAGIC after the inauguration of the second telescope, scheduled for September 21st, 2008. Multiwavelength studies are a key for the study of emission mechanisms from galactic and extragalactic sources, and Very-High Energy photon detectors are becoming crucial as the GLAST era approaches.
\end{abstract}

\section{Introduction}
The Major  Atmospheric Gamma Imaging Cherenkov (MAGIC) telescope \cite{site} is located on the Canary Island of La Palma
 and is  the largest
Imaging Atmospheric Cherenkov Telescope (IACT) worldwide.  The accessible energy range
spans from 40-50 GeV (trigger threshold at small zenith angles) up to tens of
TeV; the present analysis threshold at zenith is 60 GeV. The $5\sigma$ sensitivity of MAGIC is $\sim$~2.4\% of the Crab Nebula flux in 50
hours of observations. The relative energy resolution is about 25\% above 100 GeV and
about 20\% above 200 GeV.  The $\gamma$ point spread function (PSF) is slightly less than
$0.1$~degrees \cite{crab}. 

MAGIC has been
designed~\cite{Barrio} with the aim to achieve the lowest possible gamma energy threshold with a
ground-based gamma IACT, to fill the gap between satellite-based and ground-based gamma detectors; it is thus an ideal partner for multiwavelength (MW) campaigns.

\begin{figure}
  \begin{center}
    \epsfig{file=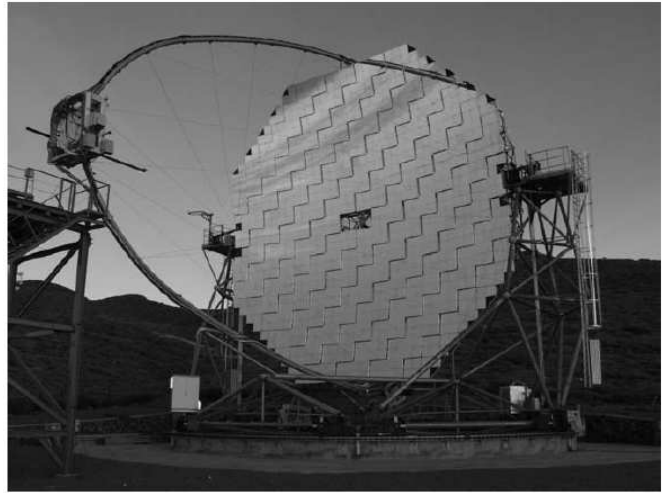,width=\columnwidth}
    \caption{The MAGIC telescope in August 2004. \label{detfin}}
  \end{center} 
\end{figure}

The observation program of MAGIC includes several galactic and extragalactic
types of sources such as Supernova Remnants (SNR), pulsars, microquasars and Active
Galactic Nuclei (AGNs). The low energy threshold allows MAGIC to extend the
observation of extragalactic sources up to z$\sim$1 and beyond. Another unique feature of MAGIC is the fast repositioning time of the telescope
that allows to observe gamma ray bursts within 25~s after the alert
by satellite detectors. Besides, MAGIC has a huge potential for studies related to fundamental physics 
(search for dark matter, study of anomalous dispersion relations for photons such as predicted by quantum gravity inspired models and by models in which the Lorentz symmetry is violated).

\section{MAGIC as it is}

MAGIC's  official observation cycles started in February 2005; each cycle spans 13 moon periods,
and is announced publicly in the MAGIC homepage. About 1/4 of the observation time (not counting that
devoted to Crab nebula technical observations) was devoted to galactic
objects, and half of the total time to Active Galactic Nuclei. 
After subtracting the bad weather and the technical runs, the MAGIC data taking efficiency is larger than 90\%;
the fraction of moon time used for data taking is close to 1/3.

MAGIC incorporates many technological innovations in order to fulfill the
requirements imposed by the physics goals\cite{nim}. 
Cost and prototyping considerations led to the
decision to construct, as a first step, a single large telescope incorporating the latest
technological developments.

A short summary of the main parameters of the MAGIC telescope is shown in the following
table:
\begin{center}
\begin{tabular}{ll}
Mount type & Alt-azimuth \\
Total weight & 65 tons \\
Max repositioning time & 22s (40s in ``safe mode"')\\
Reflector diameter, area, shape & 17m, 236m$^2$, parabolic\\
Mirror size & 0.5 $\times$ 0.5 m$^2$\\
 & 1 $\times$ 1 m$^2$ for new mirrors\\
Optics & F/D = 1\\
Mean reflectivity & 85\% \\
Pixel sizes & 397 central pixels of 1'' \\
 & 180 outer pixels of 2'' \\
 PMTs & 6 dynode\\
 DAQ rate & 2 Gsample/s \\
 DAQ rate capability & 1 kHz\\
 DAQ typical rate & 300 Hz\\
 Field of View & 3.6$^\circ$ \\
  Average PSF & 0.09$^\circ$ \\
  Average $\Delta E/E$ & 0.23\\
 Sensitivity at 5$\sigma$ (zenith): & 0.024 Crab in 50h\\
  & (0.18 Crab in 1h)\\
Analysis threshold (zenith): & 60 GeV\\
\end{tabular}
\end{center}


The sensitivity of MAGIC as calculated from the Monte Carlo simulation (MC)  is shown together with the expected sensitivity from other gamma-ray detectors in the GeV and TeV range in Fig.~2.
\begin{figure}
\begin{center}
\includegraphics[width=\columnwidth]{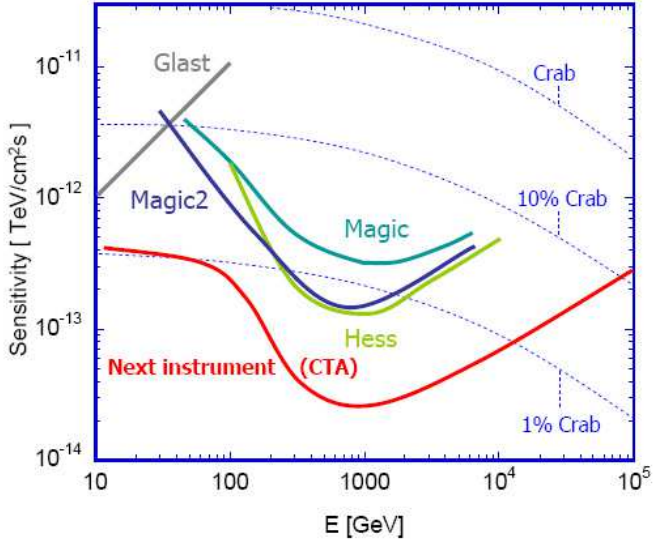}\end{center}
  \caption{Sensitivities\label{sensi} for some operating and proposed gamma detectors. The GLAST result sums 1 year of data taking, the IACTs sum 50 hours.}
\end{figure}

For calibration, image cleaning, cut optimization, and energy reconstruction,
the MAGIC standard analysis chain \cite{bretz,rwagner,sig1} is  used.  For
$\gamma$/hadron separation as well as for energy determination a
multidimensional classification technique based on the Random Forest method
\cite{breiman,rf} is used.  The cuts are  chosen such that the overall
cut efficiency for MC $\gamma$ events is about 50$\%$; cuts are established from the simulation.

In parallel with MAGIC, the
sources are usually observed with the KVA 35~cm telescope (http://tur3.tur.iac.es/),
also located on La Palma.

\section{Rules for MW with MAGIC}

MW observations in MAGIC are handled through ``standard"' and ``special" procedures.
In addition, the spokesperson has a limited (about 10\%) director-discretionary time. Up to now most of the successful MAGIC MW campaign took place through a special procedure, and as a consequence of Targets of Opportunity!

\subsection{Standard procedure}

Every year, MAGIC publishes an Announcement of Opportunity open to Guest Observations (for about 10\% of the time) and MW campaigns. The total observation time per cycle is about 1100 "`dark"' hours, plus 300 hours with moon (which implies a lower sensitivity and stricter observational constraints). External proposals are
discussed in the MAGIC Physics Committee, which tries to identify an internal PI;
the internal PI should interact with the external PI in order to guarantee that the proposal contains all information needed for a correct evaluation.
A suggestion for the internal PI could be attached to the proposal (recommended), but
people from the collaboration can volunteer to be PI.
Such simultaneous observations are in general discussed case-by-case, but special agreements exist with AGILE and are in progress with GLAST.

If the project is approved by the Time Allocation Committee, the internal PI sets up a team which helps the group in the data analysis, in order that such analysis is finished within six months and complies the quality standard of the experiment.

In any case, the first publication of data in a refereed journal should include the names of all members of the MAGIC Collaboration. In the following MW papers, from the MAGIC side, only the internal PI and the few people involved in the analysis might sign.

Co-authorships of MAGIC papers from external authors are in any case encouraged.

\subsection{Special procedure}

In case of urgent measurements (e.g., rare events, time slots from satellites) a request can arrive to any MAGIC member, who will forward it to the Physics Coordinator, or to the Spokesperson. If the urgency is recognized, the request will be communicated to the members of the MAGIC Collaboration Board and a team of MAGIC members guaranteeing a fast analysis will be set up.

As before, in any case, the first publication of data in a refereed journal should include the names of all members of MAGIC, and co-authorship is encouraged.

\section{Some results}

MW campaigns are significant both for the understanding of the emission mechanisms from astrophysical sources, since the significant features of the spectra span from the optical region to the TeV, and for the detection of new sources, since many of them can be detected only during flares, due to low fluxes. The latter aspect is especially relevant for extragalactic sources like Active Galactic Nuclei (AGN), and it has been the key for recent discoveries and observations of very distant AGN by MAGIC. MAGIC is a very good partner for MW observations due to the lowest threshold among gamma experiments, and to the best repositioning time.

\subsection{Galactic sources}

Many MW galactic campaigns are taking place, and in particular for the binary 
system LS I +61 303~\cite{lsi}, for which the VHE emission has been discovered by MAGIC. Results are expected soon.

\subsection{AGN, in particular the very distant ones}

MAGIC published up to now (August 2007) the detection of 10 extragalactic sources and all of them are well-established Active Galactic Nuclei (AGN).  The list, with increasing redshift value, is: Mkn~421 $(z=0.030)$ \cite{Mkn421}, Mkn~501 $(z=0.034)$ \cite{Mkn501}, 1ES~2344+514 $(z=0.044)$ \cite{1es2344}, Mkn~180 $(z=0.045)$ \cite{Mkn180}, 1ES~1959+650 $(z=0.047)$ \cite{1es1959}, BL Lacertae $(z=0.069)$ \cite{bllac}, 1ES~1218+304 $(z=0.182)$ \cite{1es1218}, PG~1553+113 \cite{pg1553}(unknown redshift, $0.42>z>0.09$ \cite{Sbarufatti} \cite{mazin}), 1ES1011+496 $(z=0.212)$ \cite{1es1011}, and 3C279 $(z=0.536)$ \cite{mexico}.  Here we will highlight on four recent results, obtained in a ToO program and related to the two most distant sources ever detected. MAGIC is also presently involved in multiwavelength observations of other distant sources such as 3C454.3.

MAGIC monitors some sources as Mkn~421 and Mkn~501 in order to give, possibly, alerts on exceptional activities in the VHE band.

\subsubsection{Mkn501 flare}

Together with Mkn~421, Mkn~501 is one of the most studied VHE sources, in particular by MAGIC.  
We will focus here on the flare which was detected by MAGIC during the June/July '05 observation
\cite{Mkn501}. In this case (exceptional, due to our limited FoV), MAGIC itself prompted the scientific community about the big flare.
In Fig.~3 the light curve for the night of July 9 separated in different
energy bands is shown.
An unprecedentedtly short doubling time of 2-4 minutes was detected. The rapid increase in the flux level was accompanied by a hardening of the differential spectrum. 

An energy dependent time delay of the flare peak emission
can result from the dynamics of the source, such as gradual electron acceleration in the
emitting plasma.
A somewhat more speculative issue that blazar emission permits to explore is related to
non-conventional physics, such as a violation induced in the Lorentz-Poincar\'e symmetry. A more detailed time analysis of the Mkn501 flare is in \cite{ellis}.

\begin{figure}
  \includegraphics[width=\columnwidth]{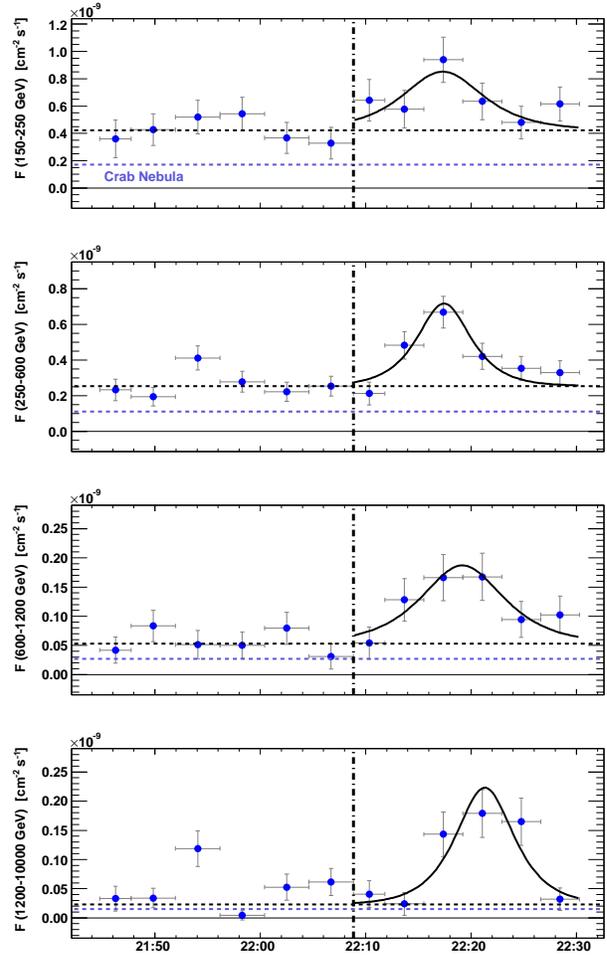}
  \caption{Mkn501 light curve for the night of July 9, 2005.}
\end{figure}

\subsubsection{BL Lac}
BL Lacertae is the historical prototype of a class of powerful $\gamma$-ray emitters: the ``BL Lac objects''.  BL Lac objects are AGNs with their jet well aligned with the observer's line of sight.  This class of object is further subdivided according to where the synchrotron emission peak lies: if it is in the sub-millimeter to optical band, the objects are classified as ``Low-frequency peaked BL Lacs'' (LBLs); if it is in the UV to X-ray band, they are referred at as ``High-frequency peaked BL Lacs'' (HBLs).

BL Lacertae was observed for 22 hours from August to December 2005 \cite{bllac}, thanks to an optical alert~\cite{gino}.  The observation showed at a level of $5.1\sigma$ a new VHE source of flux 3\% CU above 200 GeV.  The spectrum is compatible with a pure power law of index $3.6\pm0.5$.  The SED is shown in Fig.~4 together with the results from other experiments. No significant variations of the VHE flux were detected.  The source was also observed from July to September 2006 for 26 hours without any detection.  There is a remarkable agreement of the observed trend with the optical activity of the source, that, together with the discovery of Mkn 180, supports the idea of a connection (not in general simultaneous) between optical activity and increased VHE emission.

BL Lac is the first member of LBL ever detected to emit in the VHE region.  Given the very hard spectrum of the source, in agreement with LBL modeling, the low energy of MAGIC was necessary for its discovery.

\begin{figure}
  \includegraphics[width=\columnwidth]{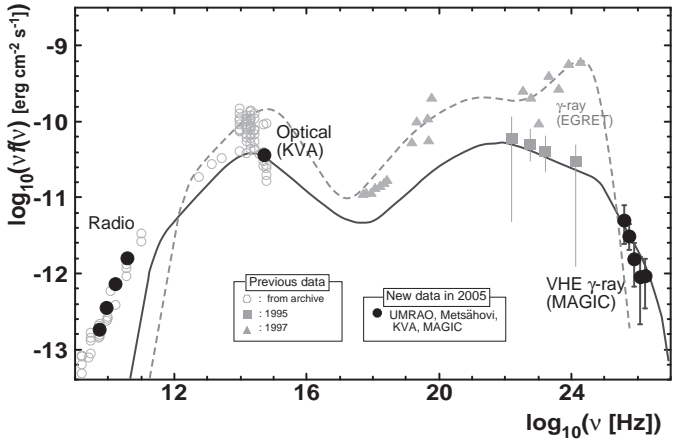}
  \caption{SED for BL Lacertae.}
\end{figure}

\subsubsection{1ES1011+496}
Very recently MAGIC  reported  the discovery of VHE $\gamma$-ray
emission from the BL Lacertae object 1ES~1011+496. The observation was
triggered by an optical outburst in March 2007 and the source was
observed with the MAGIC telescope from March to May 2007. 
Observing for 18.7 hours MAGIC found an excess of
6.2\,$\sigma$ with an integrated
flux above 200\,GeV of (1.58$\pm0.32)\times 10^{-11}$ photons cm$^{-2}$
s$^{-1}$.
The energy spectrum of 1ES~1011+496 is shown in Fig.~5.
It is well approximated by a power law with spectral index of $3.3 \pm 0.7$ after correction for 
EBL absorption.
The
redshift of 1ES~1011+496 has been detected, based on an optical spectrum that reveals the
absorption lines of the host galaxy. The redshift of $z=0.212$ makes
1ES~1011+496 the second most distant source observed to emit VHE
$\gamma$-rays up to date.

\begin{figure}
  \begin{center}
  \epsfig{figure=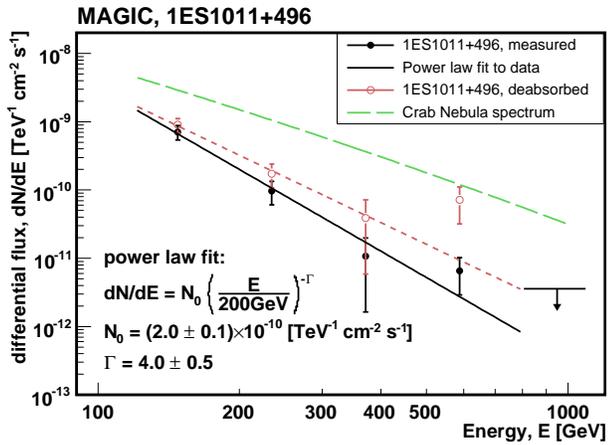,width=\columnwidth}  
  \end{center}
 \caption{The measured spectrum for 1ES1011 (black filled circles), the power-law
 fit to the data (solid line), the deabsorbed spectrum (brown
 open circles), and the fit to the deabsorbed spectrum
 (dashed brown line).}
\end{figure}

\subsubsection{3C279}

The high luminosity blazar 3C279, at $z \simeq 0.54$, had been discovered by the EGRET satellite \cite{egret}. Later, intensive simultaneous monitoring of this object was done in low energy gamma-rays,
X-rays and optical to probe crucial questions regarding the emission mechanisms and structure of jets.
However, no ground based VHE gamma ray experiment had been able to detect a signal from it due to its
high redshift and the high energy threshold of these experiments.

The MAGIC telescope observed~\cite{mexico} 3C279 between
January and March 2006, triggered by an optical
emission. Optical R-band observations were provided
by the Tuorla Observatory Blazar Monitoring
Program, with the 1.03 m telescope at the
Tuorla Observatory, Finland, and the 35 cm KVA
telescope on La Palma, Canary Island. Simultaneous WEBT observations were performed.

\begin{figure}
  \begin{center}
    \epsfig{file=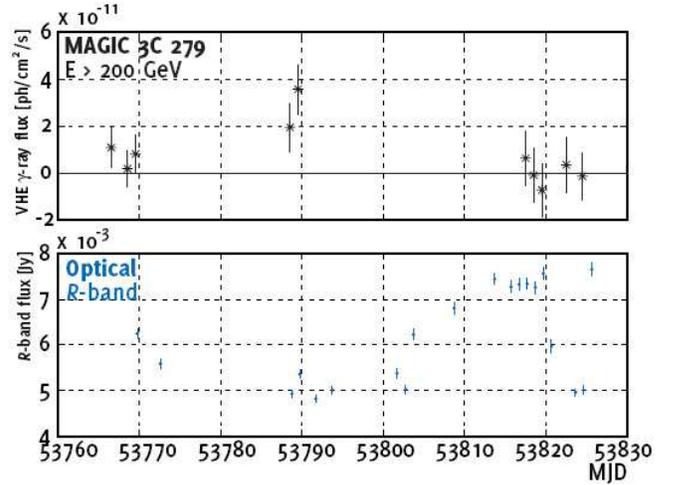,width=\columnwidth}
    \caption{MAGIC $E > 200$ GeV (top) and
optical R-band (bottom) light curves obtained for
3C279 in early 2006.
     \label{3c}}
  \end{center} 
\end{figure}

The typical observation time  by MAGIC was
around one hour per night. 
On February 22  a marginal signal
was seen; on February 23
a clear signal with an integrated
photon flux $F(E > 200$ GeV) = $(3.5 \pm 0.8) \times 10^{-11}$ photons cm$^{-2}$
s$^{-1}$ was detected. The source was observed for 62
minutes (MJD 53789.1633 to MJD 53789.2064) at zenith
angles between 35 degrees and 38 degrees. 
The MAGIC data were separated into two independent
samples: between 80 and 220 GeV
and between 220 and 600 GeV, respectively. The resulting significances are: 6.1$\sigma$ in
the low energy region and $5.1\sigma$ in the high energy
region. The detection was not accompanied by an optical
flare (Fig.~\ref{3c}) or by particularly high flux levels or outbursts
in X-rays. The MAGIC result is confirmed by three independent analyses.

\subsubsection{Special campaigns}

Finally, two additional topics.

MAGIC is linked to the GCN network for the study of GRBs, and could observe already such events simultaneously with the primary flare with SWIFT \cite{grb}.

MAGIC is also involved in a multi-messenger campaign with IceCube, hoping to detect a simultaneous VHE gamma and neutrino emission.

\section{Conclusions, and a look to the future}

Due to its low threshold and fast repositioning time, MAGIC is the ideal VHE instrument for MW observations.

The performance of the MAGIC telescope complied with the design specifications.  Since the end of commissioning, more than two full years of physics campaign followed and the data were almost completely analysed.  MAGIC detected a total of about 10 galactic and about 10 extragalactic sources, discovering new populations and new features.  Most of the new sources could be actually discovered because of the low energy threshold of MAGIC and its good sensitivity even below 100 GeV. Such a sensitivity  allowed also to resolve short term flux variability down to 2 minutes.

The construction of a second telescope, MAGIC2, close (about 80~m) to the original one, has started.  It incorporates some minor modifications suggested by the experience of running MAGIC for the last three years, as well as some  changes.
Larger (1 m$^{2}$ surface)  and lighter mirrors will be implemented, under the responsibility of the Italian INFN and INAF.
A huge improvement of the Data Acquisition System is foreseen: a new 2GHz sampler, based on a custom chip named Domino, is foreseen to reduce the Night Sky Background contribution on the signal. 
A new camera design with a total number of channels of $\sim$ 1000 is foreseen;
a large improvement is expected in the future by high quantum efficiency devices such as HPDs and and SiPMs.

MAGIC2 is expected to be ready by September 2008;  stereo observations will then be operational, allowing an increase in the sensitivity by at least a factor of 2, and other improvements in the energy and direction reconstruction.
With the advent of MAGIC2, we will reach down a level of 1\% Crab Unit in 50 hours of data taking.  Meanwhile, the 
AGILE results should come and GLAST should become fully operational, closing 
the current observational gap between $\sim1$ and $60\:\mathrm{GeV}$ and extending observations of the electromagnetic radiation, without breaks, up to almost $100\:\mathrm{TeV}$.
The inauguration of MAGIC2 is scheduled for September 21st, 2008.

\end{document}